\documentclass[letterpaper, 10 pt, journal, twoside]{IEEEtran}

\let\labelindent\relax
\usepackage{enumitem}   
\usepackage{amsmath}
\usepackage{amsthm}
\usepackage{amssymb}
\usepackage[utf8]{inputenc}
\usepackage{algorithm}
\usepackage{algorithmic}
\usepackage{hyperref}
\usepackage{subfig}

\usepackage{multirow}
 
\DeclareMathOperator*{\argmin}{arg\,min} 
 
\usepackage{centernot}
\usepackage{lipsum}
\usepackage{adjustbox}
\usepackage{pgfplots}
\usepackage{blindtext}
\usepackage{scrextend}
\usepackage{stackengine}
\usepackage{tikz}
\usetikzlibrary{
        calc,
        pgfplots.groupplots,
    }
\pgfplotsset{
    compat=1.3,
}
\pgfplotscreateplotcyclelist{my cycle list}{
    red,\\
    red, /tikz/dashed\\
    blue,\\
    blue,/tikz/dashed\\
    darkgray,\\
    darkgray, /tikz/dashed\\
}
\pgfplotscreateplotcyclelist{another list}{
    {red,dashdotted, thick},
    {blue,dashed, thick},
    {black,solid},
    {brown,dash pattern=on 5pt off 2pt,  thick},
    {teal,dashdotdotted, thick},
}

\ifCLASSINFOpdf
  
\else
  
\fi
\begin{document}
\title{
SMA-NBO: A Sequential Multi-Agent Planning with Nominal Belief-State Optimization in Target Tracking
}

\author{Tianqi Li$^{1}$, \textit{Student Member, IEEE}, Lucas W. Krakow$^2$ and Swaminathan Gopalswamy$^{1}$

\thanks{$^{1}$Tianqi Li and Swaminathan Gopalswamy are with the Department of Mechanical Engineering, 
        Texas A\&M University, College Station, TX 77840, USA
        {\tt\footnotesize \{tianqili, sgopalswamy\}@tamu.edu}}
\thanks{$^{2}$Lucas W. Krakow is with
the Bush Combat Development Complex (BCDC), Texas A\&M University, Bryan, TX 77807, USA
{\tt\footnotesize lwkrakow@tamu.edu}}
\thanks{This work has been submitted to the IEEE for possible publication. Copyright may be transferred without notice, after which this version may no longer be accessible.}
}

\markboth{Preprint Version}
{Li \MakeLowercase{\textit{et al}}: SMA-NBO: A Sequential Multi-Agent Planning with Nominal Belief-State Optimization in Target Tracking} 

\maketitle

\begin{abstract}

In target tracking with mobile multi-sensor systems, sensor deployment impacts the observation capabilities and the resulting state estimation quality. 
Based on a partially observable Markov decision process (POMDP) formulation comprised of the observable sensor dynamics, unobservable target states, and accompanying observation laws, we present a distributed information-driven solution approach to the multi-agent target tracking problem, namely, \emph{sequential multi-agent nominal belief-state optimization} (SMA-NBO). 
SMA-NBO seeks to minimize the expected tracking error via receding horizon control including a heuristic expected cost-to-go (HECTG). 
SMA-NBO incorporates a computationally efficient approximation of the target belief-state over the horizon. 
The agent-by-agent decision-making is capable of leveraging on-board (edge) compute for selecting (sub-optimal) target-tracking maneuvers exhibiting non-myopic cooperative fleet behavior. 
The optimization problem explicitly incorporates semantic information defining target occlusions from a world model. 
To illustrate the efficacy of our approach, a random occlusion forest environment is simulated. 
SMA-NBO is compared to other baseline approaches. 
The simulation results show SMA-NBO 1) maintains tracking performance and reduces the computational cost by replacing the calculation of the expected target trajectory with a single sample trajectory based on maximum a posteriori estimation; 
2) generates cooperative fleet decision by sequentially optimizing single-agent policy with efficient usage of other agents' policy of intent; 
3) aptly incorporates the multiple weighted trace penalty (MWTP) HECTG, which improves tracking performance with a computationally efficient heuristic.

\end{abstract}


\section{Introduction}

Thanks to the advances in wireless technology, we are witnessing a revolution in information acquisition driven by the convenience and advantages of wireless communications. 
A direct impact is on tracking of targets using a network of robots equipped with sensors, which has many practical applications such as in smart cities, autonomous driving, rescue tasks, surveillance monitoring, etc. 
In this context, many studies have been performed on planning for enhancing and maintaining future information gathering, such as \textit{information driven} planning  \cite{ferrari2021information}, \textit{sensor-driven control} \cite{paull2012sensor}, or \textit{active sensing acquisition} \cite{atanasov2015decentralized}.
While the concept of information-driven control is self-explanatory in its approach to improve target tracking, the realization of this approach needs to address two fundamental issues: the uncertainty of the target trajectories, and the coordination of the sensors in the network.


Information-driven control has been studied based on different target tracking objectives.
A natural and greedy objective is the maximal coverage of the sensors' field-of-view (FoV) area for richer observation \cite{cortes2004coverage}. 
However, considering the limited resources of the sensory network and the dynamic behavior of the target, maximal coverage does not guarantee the probability of detection; the Bayesian experiment design, on the other hand, plays the role of increasing such information gain from a probabilistic perspective \cite{krause2008near}. 
In information theory, entropy is one solid mathematical quantification of information gathering and is treated as the objective to minimize via observation selection and action optimization \cite{paull2012sensor, krause2008near}.
Another approach to target tracking seeks to plan the robot motion towards reduction of the mean squared error (MSE) of associated targets, which leads to the trace of covariance matrix $tr[\mathbf{P}]$ as the indicator of planning \cite{ragi2013uav, krakow2018simultaneous, ramachandran2021resilience, li2021optimizing}. 
\begin{figure*}[h!]
\centering
\captionsetup{aboveskip=-5pt}

\tikzset{every picture/.style={line width=0.75pt}} 

\begin{tikzpicture}[x=0.75pt,y=0.75pt,yscale=-1,xscale=1]

\draw (347.67,104.25) node  {\includegraphics[width=519.5pt,height=153.37pt]{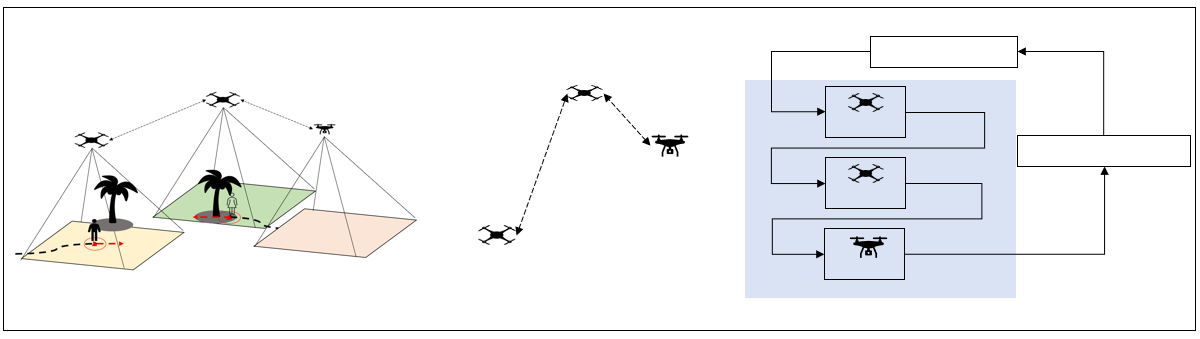}};

\draw (63.67,179.67) node [anchor=north west][inner sep=0.75pt]   [align=left] {a. Scenario};
\draw (263.33,180) node [anchor=north west][inner sep=0.75pt]   [align=left] {b. Track Fusion};
\draw (516,180) node [anchor=north west][inner sep=0.75pt]   [align=left] {c. Planning};
\draw (20,60) node [anchor=north west][inner sep=0.75pt]  [font=\scriptsize] [align=left] {UAV 1};
\draw (179,62) node [anchor=north west][inner sep=0.75pt]  [font=\scriptsize] [align=left] {UAV 3};
\draw (111,33) node [anchor=north west][inner sep=0.75pt]  [font=\scriptsize] [align=left] {UAV 2};
\draw (23,162.07) node [anchor=north west][inner sep=0.75pt]  [font=\tiny]  {$t_{1}$};
\draw (130.33,142.07) node [anchor=north west][inner sep=0.75pt]  [font=\tiny]  {$t_{2}$};
\draw (339,95) node [anchor=north west][inner sep=0.75pt]  [font=\scriptsize] [align=left] {UAV 3};
\draw (311,33) node [anchor=north west][inner sep=0.75pt]  [font=\scriptsize] [align=left] {UAV 2};
\draw (307.67,149.33) node [anchor=north west][inner sep=0.75pt]  [font=\scriptsize] [align=left] {UAV 1};
\draw (242,83.73) node [anchor=north west][inner sep=0.75pt]  [font=\scriptsize]  {$m=\{( q,\ \Omega )\}$};
\draw (591,86.73) node [anchor=north west][inner sep=0.75pt]  [font=\fontsize{0.65em}{0.78em}\selectfont]  {$\hat{\pi}^{*} \ =\left( \hat{\pi}_1^{*} \ ,\hat{\pi}_{2}^{*} \ ,\hat{\pi}_{3}^{*} \ \right)$};
\draw (525.67,28.73) node [anchor=north west][inner sep=0.75pt]  [font=\scriptsize]  {$\pi ^{p}\leftarrow \hat{\pi} ^{*}$};
\draw (484.67,73.67) node [anchor=north west][inner sep=0.75pt]  [font=\scriptsize] [align=left] {UAV 1};
\draw (484.67,116.67) node [anchor=north west][inner sep=0.75pt]  [font=\scriptsize] [align=left] {UAV 2};
\draw (486.33,160.67) node [anchor=north west][inner sep=0.75pt]  [font=\scriptsize] [align=left] {UAV 3};
\draw (528.67,53.73) node [anchor=north west][inner sep=0.75pt]  [font=\scriptsize]  {$\hat{\pi}_{1}^{*}$};
\draw (530.33,97.07) node [anchor=north west][inner sep=0.75pt]  [font=\scriptsize]  {$\hat{\pi}_{2}^{*}$};
\draw (531.33,141.73) node [anchor=north west][inner sep=0.75pt]  [font=\scriptsize]  {$\hat{\pi}_{3}^{*} \ $};

\end{tikzpicture}

\caption{The overall scenario of multi-sensor target tracking. (a) group of UAVs are tracking the OOIs (people) in an environment with occluded areas (tree shadows), the black dashed lines of targets $t_1$ and $t_2$ show the observed trajectories and the red dashed lines are nominal future trajectories for these two tracks; (b) the distributed information exchange in the team; (c)  proposed SMA information flow diagram, Algorithm \ref{alg-smanbo} sequentially generates $\hat{\pi}^{*}$ based on the intention of other agents $\pi^p$ of previous decision epoch.}
\label{fig-scenario}
\end{figure*}
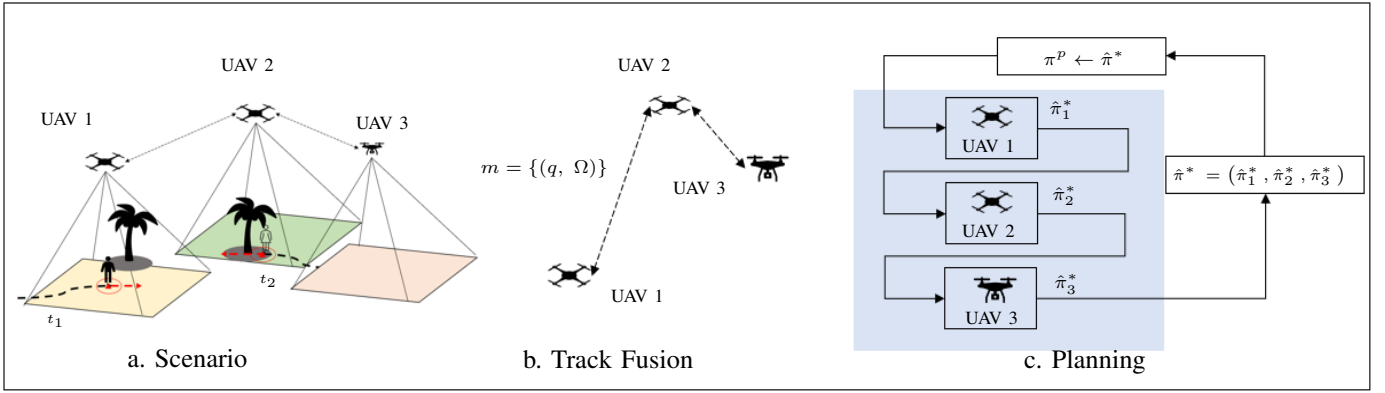

The first of the aforementioned challenges is the dynamic unknown motion of the targets. 
Tracking a target entails the planning and consequent movement of the sensors (the robots carrying the sensor) such that they attempt to maintain the targets in the FoV to generate an accurate representation of the movement of the targets. 
Since the motion of the targets is unknown apriori to the sensors, it is necessary for the planner to construct future trajectories of the targets.
Typically a  partially observable Markov decision process (POMDP) model is utilized in studying this problem, the observable states being those of the sensors and the beliefs on the states of the targets, the unobservable states being the actual states of the targets. 
In order to plan the motion of the sensors, the optimization over all possible future trajectories up to a time horizon needs to be performed. 
To make such an optimization computationally tractable, future trajectories need to be approximated. 
The Monte Carlo method is studied as one solution to such approximation, which draws target trajectory samples from beliefs of the current target state and assumptions on the dynamic behavior of the target \cite{li2021optimizing, ragi2020random}.
The disadvantage of such an approach is the computational issue for general Monte Carlo methods since a significant number of samples are required for a good approximation. 
Recently this has been addressed by generating the nominal trajectory of the targets directly from the estimated pose and dynamic behavioral models of the targets, leading to the nominal belief-state optimization (NBO) approach  \cite{ragi2013uav}. 
Our proposed approach builds on this method.

The second of the aforementioned challenges is the coordination between the sensors. 
In particular when the number of sensors $n$ increases there is a corresponding exponential growth in the action space for the control over which optimization has to be performed.
To avoid directly solving combinatorial decomposition, the greedy selection algorithm is studied with submodular objective functions \cite{corah2019distributed, friedrich2019greedy} which has $1 - e^{-1}$ best performance approximation.
\cite{ramachandran2021resilience} studies resilience to sensor deterioration by sensor reconfiguration.
The solution is a centralized mixed integer programming, and is shown to be NP-complete.
A decentralized optimization approach in Dec-POMDP is designed for mobile sensory networks without communication \cite{ragi2014decentralized}.
Even though such Dec-POMDP frees the sensors from communication load, the decentralized optimization requires each sensor to optimize the joint action, which increases the total computation.

Following the logic of the discussions above, we propose a sequential method of planning that we call SMA-NBO in this paper.
To address the challenges in multi-target tracking, SMA-NBO applies NBO as the target trajectory approximation method, optimizes agents' planning in a sequential method, and utilizes the decision of the previous epoch as other agents' intention.
Such sequential decision making in multi-agent system has a computation complexity linear in $n$ \cite{bertsekas2021multiagent}.
Specifically, the problem of target tracking by a fleet of unmanned aerial vehicles (UAV) team is addressed, using a Kalman Filter estimator based on assumptions on the dynamics of the targets, an information-driven POMDP, coordination facilitated through the SMA-NBO approach.

The primary contributions of this paper are: (i) A computationally efficient multi-agent tracking algorithm that utilizes the \textit{policy of intent} in sequential planning, extending works from \cite{li2021optimizing} that uses belief state optimization on multi-agent tracking problem and \cite{miller2009coordinated} that uses NBO on a single agent tracking problem. 
(ii) The introduction of the concept of a random forest to automatically simulate a large number of occluded environments with different quality of occlusions, and use of the random forest to statistically demonstrate the superior performance of the proposed algorithm. 
(iii) The use of a heuristic \textit{expected cost-to-go} (HECTG) that captures the expected costs beyond the fixed time used in the optimization, and demonstration of its effectiveness through the simulations in occluded environments.

\section{Problem Definition}

Consider a team of UAVs labeled by $\{1, 2, ..., n\}$ with each UAV carrying a nadir camera as its sensor, shown as Fig. \ref{fig-scenario}a.
This fleet of UAVs is tasked to track objects inside an area of interest (AOI). 
Term \textit{agent} is used in the remainder of the paper to describe the individual UAV in this robot team.
The robot team leverages communications capability to operate in a distributed manner, with each agent $i$ processing local observation $Z_{i, k}$ at time instant $k$ and generating local information, then making \textit{average-consensus} on target estimation with its neighbors in a fixed number $L$ of \textit{consensus steps}.
The data association and consensus component is consistent with the description in \cite{li2021optimizing}. 
After the agents' local processing and consensus steps, the fleet provides the estimation of target in set $\Theta_k$ at a constant sampling rate, $\Delta t$ (frequency $f_1 = 1/\Delta t$ Hz). 
The message $m$ being passed is the set of tuples of information filter, $\{(\mathbf{q, \Omega})\}$, seen in Fig. \ref{fig-scenario}b.

In this paper, we use letter $\chi$ as the true state of object of interest (OOI), $\xi$ as the mean value of estimated state of OOI with letter $t$ denotes target ID; $s$ stands for the state of UAV and $i$ is an agent ID.
We have the following assumptions in this paper.


\textit{Assumption 1}: \textit{Absolute detection}.
We assume we have a perfect detection algorithm, such that no false alarms or missed detections occur. 
Even with perfect detection, observation data is still noisy (based on common sensor models). 
Thus, estimation and data association is still required. 
Specifically we apply Joint Probabilistic Data Association which handles the target IDs, initialization and deletion of targets.

\textit{Assumption 2}: \textit{Nearly constant velocity (NCV)} motion model for targets.
Without loss of generality, the linear Kalman filter is applied in the data association algorithm with NCV motion model, i.e., for object with state $\xi_k = (p_k^x, p_k^y, v_k^x, v_k^y) \in \mathbb{R}^4$, its dynamics is 
\begin{equation}
    \label{linear-constant-velocity}
    \xi_{k+1} = \mathbf{F}_k\xi_k + \mathbf{w}_k, \mathbf{w}_k \sim \mathcal{N}(\mathbf{0},\mathbf{Q})
\end{equation}
with disturbance white noise on acceleration covariance $\sigma_a$ in $\mathbf{Q}$, and the motion model $\mathbf{F}_k$ defined as
\begin{equation}
\nonumber
\mathbf{F}_k = \begin{bmatrix}
1 & 0 & \Delta t & 0\\
0 & 1 & 0 & \Delta t\\
0 & 0 & 1 & 0 \\
0 & 0 & 0 & 1
\end{bmatrix}, 
\mathbf{Q} = \sigma_a^2\begin{bmatrix}
\frac{\Delta t^4}{4} & 0 & \frac{\Delta t^3}{2} & 0\\
0 & \frac{\Delta t^4}{4} & 0 & \frac{\Delta t^3}{2}\\
\frac{\Delta t^3}{2} & 0 & \Delta t^2 & 0 \\
0 & \frac{\Delta t^3}{2} & 0 & \Delta t^2
\end{bmatrix}
\end{equation}
Here $p_x, p_y$ and $v_x, v_y$ are the position and velocity in the $x$ and $y$ dimensions respectively. 

\textit{Assumption 3}: We assume that the appearance of our controlled agents has no impact on OOI maneuvers, i.e., $P(\chi| s) = P(\chi)$.

\textit{Assumption 4}: \textit{Fixed-time consensus on information}. 
We assume $L_0=5$ consensus time-steps during simulation for sufficient convergence \cite{battistelli2014kullback}.
Given sufficient consensus steps $L > L_0$, $\forall t \in \Theta_k$, the track estimation difference between any two agents $\forall i,j \in [n]$ is small enough, i.e., for the posterior Gaussian distribution $(\xi_{i, k|k}, \mathbf{P}_{i, k|k})$ and $(\xi_{j, k|k}, \mathbf{P}_{j, k|k})$ maintained by agent $i$ and $j$, $\exists \text{ small } \epsilon_1, \epsilon_2 \in \mathbb{R}_+, \text{s.t.} |\xi_{i, k|k} - \xi_{j, k|k}| \leq \epsilon_1$ and $|tr[\mathbf{P}_{i, k|k}] - tr[\mathbf{P}_{j, k|k}]| \leq \epsilon_2$. 

\subsection{An Information-Driven POMDP}

Similar to the work in \cite{li2021optimizing}, a POMDP model is formulated in this multi-agent scenario.
Define a POMDP model as a tuple $\mathcal{P} = (\mathcal{X, O, U, T,} C)$.

\textbf{State} $\mathcal{X}$: The state of POMDP contains the state of agents $\mathcal{S}$, state of OOIs $\chi$ and state of filter $\mathcal{F}$.
The state of all $n$ agents is defined as $\mathcal{S} = \mathcal{S}_1 \times \mathcal{S}_2 ... \times \mathcal{S}_n$, with agent $i\in[n]$ time instance $k$ the state $s_{i, k} = (p^{x}_{i, k}, p^{y}_{i, k}, \psi_{i, k}, v^{x}_{i, k}, v^{y}_{i, k})^T \in \mathbb{R}^5$ in a 2-dimensional horizontal plane including position, yaw angle and velocity.
Each agent contains its field of view (FoV) $\phi_{i, k}(s)$ as a geometric variable parametrized by its state and sensor, thus the team's FoV is described as $\phi_k(s_k) = (\phi_{1, k}, ..., \phi_{n, k})$.
Additionally, with a semantic map $W$ which contains the information of occlusion area in the AOI where no observation can be acquired by any agent, the semantic FoV is denoted as $\phi_k(s; W)$.
This FoV information $\phi_k(s; W)$ with semantic map is implicitly contained in the agent state.
OOIs' state $\chi$ contains all position and velocities of current OOIs.
The filter state $\mathcal{F}_k = (\xi_{k|k}, \mathbf{P}_{k|k})$ maintains tracks represented by Gaussian distributions with posterior mean $\xi_{k|k}$ and covariance matrix $\mathbf{P}_{k|k}$.
To avoid the redundancy of terminology, vector mean $\xi$ and covariance matrix $\mathbf{P}$ contains all targets, i.e., $\xi_{k|k} = (\xi_{1, k|k}, ..., \xi_{m_k, k|k})$ and block-diagonal matrix $\mathbf{P}_{k|k} = diag(\mathbf{P}_{1, k|k}, ..., \mathbf{P}_{m_k, k|k})$ for for total $m_k$ targets.
The dynamics of the target is based on the linear constant velocity motion \eqref{linear-constant-velocity}.
The POMDP state is summarized as $x_k = (s_k,\chi_k, \mathcal{F}_k)$.

\textbf{Observation and Observation Law} $\mathcal{O}$:
Observation data of each agent is a set of 2-dimensional positions captured as targets.
Given the state estimate of a target $t$ at time instance $k$, $\xi_{t,k} \in \mathbb{R}^4$, the observation model is 
\begin{equation}
    \label{observation-model}
    \mathbf{z}_{t, k} = \mathbf{H}_k\xi_{t, k} + \mathbf{v}_{t, k},  \mathbf{v}_{t, k} \sim \mathcal{N}(0, \mathbf{R})
\end{equation}
$\mathbf{v}_k$ represents the measurement noise. 
The \textit{range-bearing sensor} model is applied as the observation model, for agent $i$ with state $s_{i, k}$, let $r_k$ and $\rho_k$ denote the estimated range and bearing of the target $t$, $r_k = \max(d(s_{i, k}, \xi_{t, k}), r_0)$ with Euclidean distance $d(x, y) = ||x - y||$, $r_0$ is defined as the minimal effective range threshold, and $\rho_k$ is the angular measurement between the sensor and target.
The observation covariance matrix $\mathbf{R}(s_{i, k}, \xi_{t, k})$ for target $t$ is 
\begin{equation}
\label{observation-model}
\mathbf{R}(s_{i, k}, \xi_{t, k}) = \alpha_i \mathbf{G}(\rho_k)
\begin{bmatrix}
0.1r_k & 0\\
0 & 0.1\pi r_k
\end{bmatrix}
\mathbf{G}(\rho_k)^T
\end{equation}
where $\mathbf{G}(\rho)$ is the rotation matrix of angle $\rho$, and sensing quality factor $\alpha_i$ is a scalar in this uncertainty matrix that varies by agent.
The observation model \eqref{observation-model} for a given sensor can be extended for all the targets it observes to obtain the block diagonal $\mathbf{H}$ and $\mathbf{R}$ matrices for each agent.
This observation setup imposes the spatially varying measurement error \cite{krakow2018simultaneous}. 
With observation law defined, the observation of $n$ agents at each time step is $Z_k \in \mathcal{O}, Z_k = \{Z_{i, k}|i\in [n]\}$.

\textbf{Action} $\mathcal{U}$: The action of an agent $i$ is the command of horizontal velocity, i.e. $u_{i} = (u^x_i, u^y_i) \in \mathbb{R}^2$ with maximum velocity constraint $|u| \leq v_{max}$.
We assume the motion of agent is deterministic by
\begin{equation}
    \label{agent-dynamic}
    s_{i, k+1} = f(s_{i, k}, u_{i, k}) =
\begin{bmatrix}
p_{i, k}^x + u_{i, k}^x\Delta t\\
p_{i, k}^y + u_{i, k}^y\Delta t\\
arctan(u_{i, k}^y/u_{i, k}^x)\\
u_{i, k}^x\\
u_{i, k}^y
\end{bmatrix}.
\end{equation}
The joint action domain over $n$ agent is $\mathcal{U} = \mathcal{U}_1 \times ... \times \mathcal{U}_n$.

\textbf{State Transition} $\mathcal{T}$: 
State transition law is defined as the mapping $\mathcal{T}: \mathcal{X} \times \mathcal{U} \xrightarrow[]{} \mathcal{X}$.
In the state $x_k\in \mathcal{X}, x_k = (s_k, \chi_k, \mathcal{F}_k),u_k\in\mathcal{U}$, the state transition is decomposed into the following:
\begin{itemize}
    \item the agents' deterministic transitions \eqref{agent-dynamic};
    \item OOI states $\chi_{t, k} \in \chi_k$ are stochastic, independent on sensor state $s_k$ and control variable $u_k$;
    \item the filter state $\mathcal{F}_k$ is dependent on both agent state $s_k$, FoV of sensors $\phi_k(s_k; W)$, observation law $\mathcal{O}$ and target state $\chi_k$.
\end{itemize}

\textbf{Cost} $C$: The cost function is a mapping $\mathcal{X} \times \mathcal{U} \xrightarrow[]{} \mathbb{R}_{\geq 0}$, with the overall optimization objective of minimizing the total expected Mean Square Error (MSE). Correspondingly, the one-step cost for the system is 
\begin{equation}
\label{cost}
C(x_k, u_k) = 
 \mathbb{E}_{w_k, v_{k+1}}
         \bigl[ \, ||\chi_k - \xi_k||^2 \vert\, x_k, u_k \bigr]
\end{equation}

\subsection{Belief-State MDP}

Because of the partial observability in the state, the unobserved variable is depicted by the probability distribution.
Applying the MDP solution to POMDP requires the definition of belief state \cite{kaelbling1998planning}: $b_k(x) := b(x_k) \in \mathcal{B(X)}$, $b_k(x) = P_{x_k}(x| Z_0, Z_1, ..., Z_k, u_0, ..., u_k) = (b^s_k, b^\chi_k, b^\xi_k, b^P_k)$.
All belief states of observable state components ($b^s_k, b^\xi_k, b^P_k$) can be represented by Dirac delta function, e.g. $b^s_k = \delta(s - s_k)$;
the target belief state is represented by filter's posterior estimate $b^\chi_k \sim \mathcal{N} (\xi_{k|k}, \mathbf{P}_{k|k})$. 
Utilizing the belief state as the system state describes a fully observable system with information sufficiently characterized by the above defined distributions. 
Now standard MDP theory can be applied to the resulting MDP $\mathcal{M = (\mathcal{B, O, U, \Tilde{T}, }} \Tilde{C})$ which includes the adaptation of the cost $\Tilde{c}$ and state transition laws $\Tilde{\mathcal{T}}$.

\textbf{Belief-State Transition} $\Tilde{\mathcal{T}}$: 
Belief-state transition law is the mapping $\Tilde{\mathcal{T}}: \mathcal{B} \times \mathcal{U} \xrightarrow[]{} \mathcal{B}$.
Following \eqref{agent-dynamic} for each agent, the action to a belief state makes deterministic transition to agents state $s\in\mathcal{S}$; 
the filter state $\mathcal{F}$ depends on sensors FoVs $\phi_k(s_k; W)$ and OOI state $\chi_k$, which is interpreted by the state belief $b_k(x)$;
OOI state is action-independent.

\textbf{Belief Cost} $\Tilde{C}$: Based on state transition law $\Tilde{\mathcal{T}}$, action $u\in\mathcal{U}$ makes one-step cost over the belief state $b_k$ with the expected value of $\Tilde{C}(b_k, u)$.
\begin{equation}
\begin{aligned}
\Tilde{C}(b_k, u) & = \int C(x_k, u)b_k(x)dx \label{belief-costs}\\
& = \int \mathbb{E}_{w_k, v_{k+1}}
\bigl[ \, ||\chi_{k+1} - \xi_{k+1}||^2 \vert s_k, \xi_k, u
\bigr] \,b^\chi_k(x)dx \\
& = tr[\mathbf{P}_{k+1|k+1}]
\end{aligned}
\end{equation}
The third equality is derived based on the Gaussian noise assumption \cite{miller2009coordinated}.

Consistent with previous work \cite{li2021optimizing}, we seek a solution from receding horizon control over fixed time horizon $H$ in belief-state MDP defined.
Denote the policy over belief state as $\pi: \mathcal{B} \xrightarrow[]{} \mathcal{U}$, we define the cumulative cost over horizon $H$ as 
\begin{equation}
\label{J_H}
J_H^\pi(b_k) = \mathbb{E}\biggl[ \,\sum_{l=k}^{k+H-1} \Tilde{C}(b_l, \pi(b_l)) \bigg\vert b_k \,\biggl] 
\end{equation}
then the optimal policy $\pi^*$ follows the Bellman's principle
\begin{equation}
    \pi^*(b_k) \in \argmin_u (\Tilde{C}(b_k, u) + \mathbb{E}[J_{H-1}^*(b_{k+1})|b_k, u]) 
\end{equation}
and $J^*_{H-1}$ is the minimized objective cost over subsequent horizon $H-1$.
Define the \textit{Q-value} function over horizon $H$ as $Q_H: \mathcal{B} \times \mathcal{U} \xrightarrow[]{} \mathbb{R}$, which is the expected cost of executing $u$ in belief state $b_0$
\begin{equation}
    \label{Q-value}
    Q_{H}(b_k, u) = \Tilde{C}(b_k, u) + \mathbb{E}[J^*_{H-1}(b_{k+1})|b_k, u]
\end{equation}
and optimal policy $\pi^*(b_k) \in \argmin_{u\in\mathcal{U}} Q_H(b_k, u)$.

\section{SMA-NBO}

\begin{algorithm}[t!]
\caption{Nominal Belief Optimization}
\label{alg-nbo}
\begin{algorithmic} 
\REQUIRE Initial belief state $b_k = (b^s_k, b^\chi_k, b^\xi_k, b^P_k)$, target mean $\hat{\xi}_k = \xi_k$.\\
\STATE 1. Generate nominal trajectory $\{\Hat{b}^\chi\}_{k+1:k+H}$
\FOR{$l$ in $k+1$ to $k+H$}
\STATE $\hat{\xi}_l = \mathbf{F}_{l-1} \hat{\xi}_{l-1}$
\STATE $\hat{b}^\chi_l = \delta(\chi - \hat{\xi}_l)$
\ENDFOR
\STATE 2. Policy optimization
\STATE Optimize $\Tilde{J}^\pi_H(b_k)$ by policy $\pi^*$.
\RETURN $\pi^*$
\end{algorithmic}
\end{algorithm}

Given the problem setup, we introduce our SMA-NBO as the planning algorithm of multi-sensor target tracking.
Since the non-fixed number of targets and redundant filter state dimension, methods like Q-learning that directly explores the whole state $\mathcal{B}$ are unrealistic in computation.
Our algorithm SMA-NBO seeks the proper approximation of optimization in \eqref{Q-value} from two aspects: given current belief state $b_0$, the approximation of future belief state $b_k$ (NBO) and joint action optimization in the multi-agent team (SMA).

\subsection{Nominal Belief Optimization}
To obtain the optimal policy $\pi^*(b_k)$ given belief state $b_k$, the subsequent \textit{expected cost-to-go} (ECTG) $J^*_{H-1}$ is essential in \eqref{Q-value}.
Denote the future belief of target trajectory over horizon $H$ as $\{b\}_{k+1:k+H}$.
However, the future target trajectory belief $\{b^\chi\}_{k+1:k+H}$ is stochastic and independent from action selection by Assumption 2.
The approach of NBO is introduced for action optimization by approximation to future target belief $\{\Hat{b}^\chi\}_{k+1:k+H}$. 

Algorithm \ref{alg-nbo} is a description of NBO with linear dynamic target assumption \eqref{linear-constant-velocity}.
Step 1 generates the nominal trajectory, which is maximum a posterior (MAP) estimate of the belief-state distribution by ignoring the disturbance term $\mathbf{w}_k$ in the NCV transition law.
The approximate cost over horizon is
\begin{equation}
\label{V-nbo}
\Tilde{J}_H^\pi(b_k) = E\biggl[ \,\sum_{l=k}^{k+H-1} \Tilde{C}(\hat{b}_l, \pi(\hat{b}_l)) \bigg\vert b_k, \hat{b}^\chi \,\biggl] 
\end{equation}
In Step 2, the action optimization is obtained from approximation function $\tilde{J}_H^\pi$.
The output of NBO $\pi^*(b_k) = (u_k, ..., u_{k+H-1})$ is the joint fleet plan for $H$ future steps, and the agents only execute the first action $u_k$ as their immediate control command.
The approximate objective $\Tilde{J}_H^\pi$ is typically called truncated version since it only calculates the cost over the fixed-length horizon \cite{bertsekas2021multiagent, miller2009coordinated}. 

Receding horizon control, optimizing actions over a fixed horizon of $H$, garners computational savings by considering only the initial portion of the infinite horizon. 
Additionally, the accuracy of the nominal trajectory approximation decreases with the extended horizon lengths due to compounding stochasticity resulting from the target motion model. 
In contrast, accounting for impacts of actions beyond horizon $H$ in the planning objective further emphasizes the non-myopic behavior of the multi-agent system.  
For this reason, we implement a HECTG term $\hat{J}(\hat{b}_{k+H})$ added at the end of horizon. 
The HECTG not only emphasizes non-myopic planning but also avoids the detrimental computational impacts of over-extended horizons.
In this way we better approximate the infinite horizon control with NBO 
\begin{equation}
    \label{infi-obj}
    \Tilde{J}_\infty^\pi(b_k) = E\biggl[ \,\sum_{l=k}^{k+H-1} \Tilde{C}(\hat{b}_l, \pi(\hat{b}_l)) + \hat{J}(\hat{b}_{k+H}) \bigg\vert b_k, \hat{b}^\chi \,\biggl] 
\end{equation}

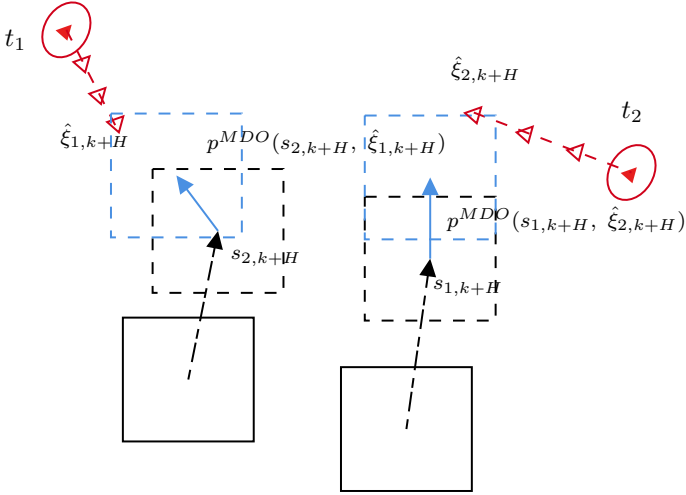
\begin{figure}
\tikzset{every picture/.style={line width=0.75pt}} 
\centering
\captionsetup{aboveskip=5pt}

\begin{tikzpicture}[x=0.75pt,y=0.75pt,yscale=-1,xscale=1]

\draw  [draw opacity=0][fill={rgb, 255:red, 230; green, 30; blue, 30 }  ,fill opacity=1 ] (46.21,50.1) -- (43.99,58.2) -- (38.06,52.99) -- cycle ;
\draw  [color={rgb, 255:red, 208; green, 2; blue, 27 }  ,draw opacity=1 ] (34.14,47.01) .. controls (38.39,40.07) and (46.04,37.01) .. (51.22,40.18) .. controls (56.4,43.35) and (57.15,51.56) .. (52.9,58.5) .. controls (48.64,65.45) and (40.99,68.51) .. (35.81,65.34) .. controls (30.63,62.16) and (29.88,53.96) .. (34.14,47.01) -- cycle ;
\draw   (72,198) -- (138,198) -- (138,260.5) -- (72,260.5) -- cycle ;
\draw  [dash pattern={on 4.5pt off 4.5pt}] (87,123) -- (153,123) -- (153,185.5) -- (87,185.5) -- cycle ;
\draw  [draw opacity=0][fill={rgb, 255:red, 230; green, 30; blue, 30 }  ,fill opacity=1 ] (331.21,122.1) -- (328.99,130.2) -- (323.06,124.99) -- cycle ;
\draw  [color={rgb, 255:red, 208; green, 2; blue, 27 }  ,draw opacity=1 ] (319.14,119.01) .. controls (323.39,112.07) and (331.04,109.01) .. (336.22,112.18) .. controls (341.4,115.35) and (342.15,123.56) .. (337.9,130.5) .. controls (333.64,137.45) and (325.99,140.51) .. (320.81,137.34) .. controls (315.63,134.16) and (314.88,125.96) .. (319.14,119.01) -- cycle ;
\draw  [dash pattern={on 3.75pt off 3pt on 7.5pt off 1.5pt}]  (105,229.25) -- (119.41,157.19) ;
\draw [shift={(120,154.25)}, rotate = 101.31] [fill={rgb, 255:red, 0; green, 0; blue, 0 }  ][line width=0.08]  [draw opacity=0] (8.93,-4.29) -- (0,0) -- (8.93,4.29) -- cycle    ;
\draw   (182,223) -- (248,223) -- (248,285.5) -- (182,285.5) -- cycle ;
\draw  [dash pattern={on 4.5pt off 4.5pt}] (194,137) -- (260,137) -- (260,199.5) -- (194,199.5) -- cycle ;
\draw  [dash pattern={on 3.75pt off 3pt on 7.5pt off 1.5pt}]  (215,254.25) -- (226.59,171.22) ;
\draw [shift={(227,168.25)}, rotate = 97.94] [fill={rgb, 255:red, 0; green, 0; blue, 0 }  ][line width=0.08]  [draw opacity=0] (8.93,-4.29) -- (0,0) -- (8.93,4.29) -- cycle    ;
\draw [color={rgb, 255:red, 208; green, 2; blue, 27 }  ,draw opacity=1 ] [dash pattern={on 4.5pt off 4.5pt}]  (43.52,52.76) -- (70,104.5) ;
\draw  [color={rgb, 255:red, 208; green, 2; blue, 27 }  ,draw opacity=1 ] (55.21,66.1) -- (52.99,74.2) -- (47.06,68.99) -- cycle ;
\draw  [color={rgb, 255:red, 208; green, 2; blue, 27 }  ,draw opacity=1 ] (63.21,82.1) -- (60.99,90.2) -- (55.06,84.99) -- cycle ;
\draw  [color={rgb, 255:red, 208; green, 2; blue, 27 }  ,draw opacity=1 ] (72.22,96.4) -- (70,104.5) -- (64.06,99.28) -- cycle ;
\draw  [color={rgb, 255:red, 74; green, 144; blue, 226 }  ,draw opacity=1 ][dash pattern={on 4.5pt off 4.5pt}] (66,95) -- (132,95) -- (132,157.5) -- (66,157.5) -- cycle ;
\draw [color={rgb, 255:red, 74; green, 144; blue, 226 }  ,draw opacity=1 ]   (120,154.25) -- (100.8,128.65) ;
\draw [shift={(99,126.25)}, rotate = 53.13] [fill={rgb, 255:red, 74; green, 144; blue, 226 }  ,fill opacity=1 ][line width=0.08]  [draw opacity=0] (8.93,-4.29) -- (0,0) -- (8.93,4.29) -- cycle    ;
\draw [color={rgb, 255:red, 208; green, 2; blue, 27 }  ,draw opacity=1 ] [dash pattern={on 4.5pt off 4.5pt}]  (249,94.5) -- (328.52,124.76) ;
\draw  [color={rgb, 255:red, 74; green, 144; blue, 226 }  ,draw opacity=1 ][dash pattern={on 4.5pt off 4.5pt}] (194,96) -- (260,96) -- (260,158.5) -- (194,158.5) -- cycle ;
\draw [color={rgb, 255:red, 74; green, 144; blue, 226 }  ,draw opacity=1 ]   (227,168.25) -- (227,130.25) ;
\draw [shift={(227,127.25)}, rotate = 90] [fill={rgb, 255:red, 74; green, 144; blue, 226 }  ,fill opacity=1 ][line width=0.08]  [draw opacity=0] (8.93,-4.29) -- (0,0) -- (8.93,4.29) -- cycle    ;
\draw  [color={rgb, 255:red, 208; green, 2; blue, 27 }  ,draw opacity=1 ] (278.71,101.6) -- (276.49,109.7) -- (270.56,104.49) -- cycle ;
\draw  [color={rgb, 255:red, 208; green, 2; blue, 27 }  ,draw opacity=1 ] (304.21,110.6) -- (301.99,118.7) -- (296.06,113.49) -- cycle ;
\draw  [color={rgb, 255:red, 208; green, 2; blue, 27 }  ,draw opacity=1 ] (252.7,91.85) -- (250.48,99.94) -- (244.54,94.73) -- cycle ;

\draw (11,51.4) node [anchor=north west][inner sep=0.75pt]    {$t_{1}$};
\draw (322,87.4) node [anchor=north west][inner sep=0.75pt]    {$t_{2}$};
\draw (39,98.4) node [anchor=north west][inner sep=0.75pt]  [font=\footnotesize]  {$\hat{\xi }_{1,k+H}$};
\draw (236,64.4) node [anchor=north west][inner sep=0.75pt]  [font=\footnotesize]  {$\hat{\xi }_{2,k+H}$};
\draw (227,177.4) node [anchor=north west][inner sep=0.75pt]  [font=\footnotesize]  {$s_{1,k+H}$};
\draw (125,162.4) node [anchor=north west][inner sep=0.75pt]  [font=\footnotesize]  {$s_{2,k+H}$};
\draw (113,101.4) node [anchor=north west][inner sep=0.75pt]  [font=\footnotesize]  {$p^{MDO}( s_{2,k+H} ,\ \hat{\xi }_{1,k+H})$};
\draw (234,140.4) node [anchor=north west][inner sep=0.75pt]  [font=\footnotesize]  {$p^{MDO}( s_{1,k+H} ,\ \hat{\xi }_{2,k+H})$};

\end{tikzpicture}

\caption{An example of MWTP. 
Sensors' current FoVs positions are rectangles of solid line, and the targets in set $T$ defined are red solid triangles and ellipses as means and covariances. 
The nominal trajectories of targets in the horizon ($H=3$ in this case) are dashed lines with mean in hollow triangles.
Rectangles of black dashed line are the final positions of sensors FoVs, and rectangles of blue dashed line are the adjusted positions of FoVs $p^{MDO}(s_{i, k+H}, \hat{\xi}_{t, k+H})$ when matching target $t$ to sensor $i$.}
\label{fig-mwtp}
\end{figure}

For multi-sensor multi-target tracking, \cite{miller2009coordinated} proposed MWTP as such a HECTG. 
We have adapted MWTP for FoV considerations.
Denote the set of targets at end of planning horizon that are outside sensors FoVs as set $T = \{\hat{\xi}_{t, k+H}: \hat{\xi}_{t, k+H} \notin \phi(\hat{b}^{s}_{k+H})\}$, and sensors set at end of horizon as $S = \{\hat{b}^{s_i}_{k+H}\}$. 
MWTP is detailed in Algorithm \ref{alg-mwtp}, which takes the targets with higher uncertainty ($tr[\mathbf{P}_{t, k+H}]$) to match the nearest sensors first.
The if statement enforces a bipartite matching such that each sensor accrues a penalty based on at most one unobservable target.
In Step 2 of Algorithm \ref{alg-mwtp}, the matching of targets to sensors will generate the adjusted position $p^{MDO}(s_{i, k+H}, \hat{\xi}_{t, k+H})$, which is the position of minimum distance to observation (MDO) for sensor $i$ that its FoV $\phi(s_{i, k+H})$ just covers the targets nominal mean position $\hat{\xi}_{t, k+H}$ with minimal displacement.
An example is shown in Fig. \ref{fig-mwtp}, targets not covered in sensors FoVs are in set $T = \{t_1, t_2\}$, and we have sensors' positions at end of horizon $S = \{s_{1, k+H}, s_{2, k+H}\}$.
The MWTP matches sensor 1 to target 2, and sensor 2 to target 1.

\subsection{Sequential Multi-Agent Decision Making}

\begin{algorithm}[t!]
\caption{Multiple Weighted Trace Penalty}
\label{alg-mwtp}
\begin{algorithmic} 
\REQUIRE Targets outside sensors FoVs at end of horizon $T$, sensors set at end of horizon $S$.\\
\STATE 1. Sort targets $T$ in decreasing order of $tr[\hat{\mathbf{P}}_{t, k+H}]$
\STATE 2. Initialize $\hat{J} = 0$ and all sensors $\forall i\in[n], D_i = 0$
\FOR{target $t$ in $T$}
\STATE Get sensor $i \in \argmin_{i\in[n]} D_i + d(s_{i, k+H}, \hat{\xi}_{t, k+H})$
\IF{$D_i = 0$}
\STATE $\hat{J} = \hat{J} + \beta d(s_{i, k+H}, \hat{\xi}_{t, k+H}) tr[\mathbf{\hat{P}}_{t, k+H}]$
\ENDIF
\STATE $D_i = D_i + d(s_{i, k+H}, \hat{\xi}_{t, k+H})$
\STATE $s_{i, k+H} = p^{MDO}(s_{i, k+H}, \hat{\xi}_{t, k+H})$
\ENDFOR
\RETURN $\hat{J}$
\end{algorithmic}
\end{algorithm}

In the multi-agent system, the optimization problem is typically constrained by the resource limit, for example, the exponential complexity growth with the number of agents in the team.
Such issue exists in NBO which defines the joint action space as $\mathcal{U}$ and optimizes the team's policy $\pi^*$ in Step 2 of Algorithm \ref{alg-nbo}.
An alternative but suboptimal way is to consider a sequence in optimizing single agents' actions based on the intention of other agents. 
We denote the \textit{policy of intent} as $\bar{\pi}$.
Define the policy of agent $i$ as $\pi_i(b_k) = (u_{i, k}, ..., u_{i, k + H-1})$ (i.e., the sequence of actions over the planning horizon) and the policy for a sequence of agents from $i$ to $i+j$ as $\pi_{i:i+j} = (\pi_i,\pi_{i+1} ..., \pi_{i+j})$.
Then at each decision epoch, the action optimization starts with first agent by
\begin{equation}
    \label{sma-1}
    \hat{\pi}_1^{*} \in \argmin_{\pi_1} J_H^{(\pi_1, \bar{\pi}_{2:n})}(b_k)
\end{equation}
This decision is passed to the next agent in the  optimization sequence. 
The general optimization of agent $i$ is
\begin{equation}
    \label{sma-i}
    \hat{\pi}_i^{*} \in \argmin_{\pi_i} J_H^{(\hat{\pi}_{1:i-1}^{*}, \pi_i, \bar{\pi}_{i+1:n})}(b_k)
\end{equation}

However, the policy of intent $\bar{\pi}$ is a key factor in optimization.
After agent $j$ executes action $u_{j,k}$ at time $k$, there is a remaining action sequence $\pi^p_j = ( u_{j,k+1},...u_{j,k+H})$. 
This remainder provides insight to the $j$th agent's \textit{future intentions} and can be used at time $k+1$ to inform the control decisions of agents $1:i$ in \eqref{sma-i} s.t. $i<j$ by simply extending $\pi^p_j$ with an additional action $\bar{u}_j$ generated by a heuristic base single-agent base policy. 
This yields an approximate $H$-step policy $\bar{\pi}_j = (\pi^p_j, \bar{u}_j)$ containing agent $j$'s future intent.


Fig. \ref{fig-scenario}c shows the proposed SMA in block diagram and Algorithm \ref{alg-smanbo} details the SMA-NBO.
The reduction of computation is the main advantage of SMA-NBO.
For the general searching method, the time complexity of the optimization in \eqref{sma-i} is the planning domain of agent $i$ over horizon $H$, denoted as $O(|\mathcal{U}_i^H|)$, and SMA-NBO has complexity increase linear to agent number, $O(n|\mathcal{U}_i^H|)$;
while for joint optimization of agents the complexity grows exponentially, $O(|\mathcal{U}_i^H|^n)$.
Also, such sequential decision-making makes distributed computations possible rather than a single agent making and commanding fleet-wide decisions.
\begin{algorithm}[t!]
\caption{SMA-NBO}
\label{alg-smanbo}
\begin{algorithmic} 
\REQUIRE Initial belief state $b_k = (b^s_k, b^\chi_k, b^\xi_k, b^P_k)$, target mean $\hat{\xi}_k = \xi_k$, previous decision $\pi^p$\\
\STATE 1. Generate nominal trajectory $\{\Hat{b}^\chi\}_{k+1:k+H}$
\\
\STATE 2. Generate policy of intent $\forall i \in [n], \bar{\pi}_i = (\pi^p_i, \bar{u}_i)$
\STATE 3. Sequential multi-agent decision making
\FOR{agent $i$ from 1 to $n$}
\STATE $\hat{\pi}_i^{*} \in \argmin_{\pi_i} \tilde{J}_H^{(\hat{\pi}_{1:i-1}^{*}, \pi_i, \bar{\pi}_{i+1:n})}(b_k)$
\ENDFOR
\RETURN $\hat{\pi}^{*}$;

\end{algorithmic}
\end{algorithm}

\begin{figure}
    \centering
    \captionsetup{aboveskip=-5pt}

\tikzset{every picture/.style={line width=0.75pt}} 

\begin{tikzpicture}[x=0.75pt,y=0.75pt,yscale=-1,xscale=1]

\draw (161.13,114.75) node  {\includegraphics[width = \columnwidth]{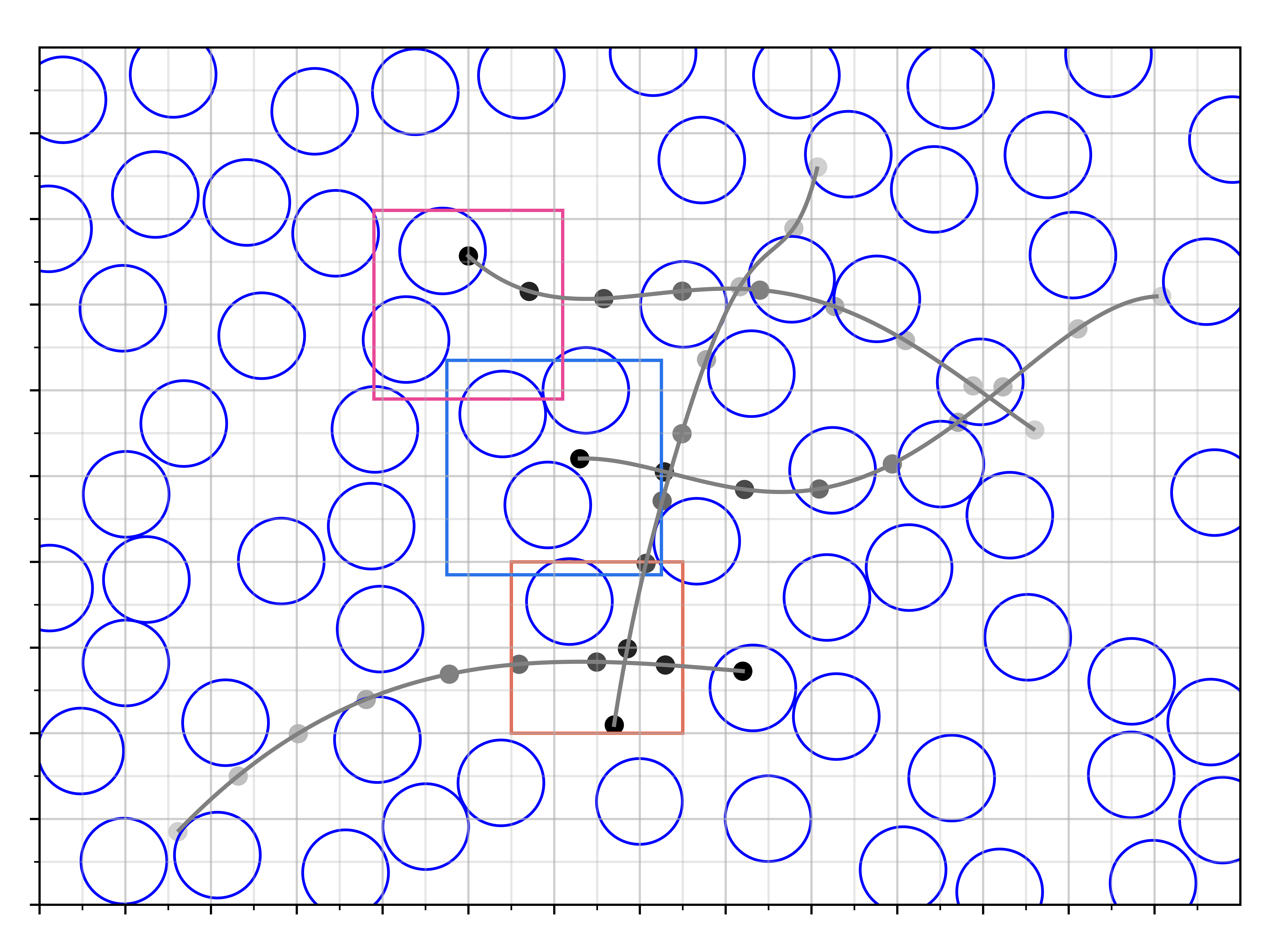}};

\draw (72.5,69) node [anchor=north west][inner sep=0.75pt]  [font=\tiny,color={rgb, 255:red, 204; green, 6; blue, 103 }  ,opacity=1 ] [align=left] {\textbf{UAV 3}};
\draw (89.25,125.25) node [anchor=north west][inner sep=0.75pt]  [font=\tiny,color={rgb, 255:red, 7; green, 95; blue, 237 }  ,opacity=1 ] [align=left] {\textbf{UAV 2}};
\draw (105.75,165) node [anchor=north west][inner sep=0.75pt]  [font=\tiny,color={rgb, 255:red, 204; green, 6; blue, 103 }  ,opacity=1 ] [align=left] {\textcolor[rgb]{0.94,0.23,0.09}{\textbf{UAV 1}}};
\draw (146.25,181.75) node [anchor=north west][inner sep=0.75pt]  [font=\tiny,color={rgb, 255:red, 204; green, 6; blue, 103 }  ,opacity=1 ] [align=left] {\textcolor[rgb]{0.01,0.01,0.01}{Target 1}};
\draw (203.5,155.5) node [anchor=north west][inner sep=0.75pt]  [font=\tiny,color={rgb, 255:red, 204; green, 6; blue, 103 }  ,opacity=1 ] [align=left] {\textcolor[rgb]{0.01,0.01,0.01}{Target 2}};
\draw (125.5,118.5) node [anchor=north west][inner sep=0.75pt]  [font=\tiny,color={rgb, 255:red, 204; green, 6; blue, 103 }  ,opacity=1 ] [align=left] {\textcolor[rgb]{0.01,0.01,0.01}{Target 3}};
\draw (121.5,39.25) node [anchor=north west][inner sep=0.75pt]  [font=\tiny,color={rgb, 255:red, 204; green, 6; blue, 103 }  ,opacity=1 ] [align=left] {\textcolor[rgb]{0.01,0.01,0.01}{Target 4}};

\end{tikzpicture}

    \caption{Multi-sensor multi-target tracking environment, which contains 3 UAVs with their FoVs in rectangles, 4 targets and trajectory crossovers and divergence.}
    \label{fig-simenv}
\end{figure}
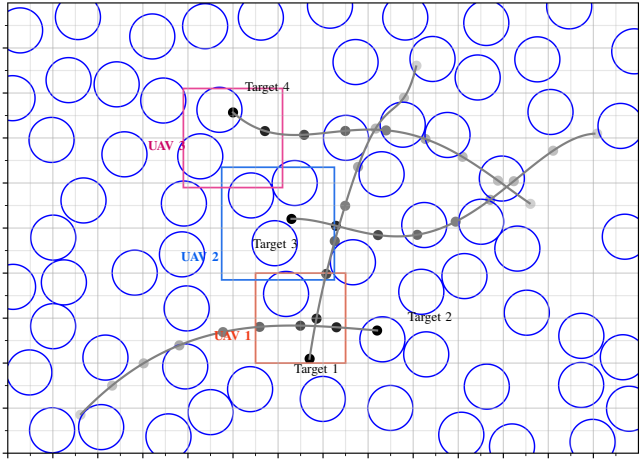

\section{Simulation Results}

This section presents the simulation setup, performance results and conclusions drawn from the SMA-NBO experiments.
Fig. \ref{fig-simenv} shows the general simulation environment of a 3-heterogeneous UAV tracking team.
These UAVs are detecting people as targets in a $150\times 100$ m rectangle AOI with sensing frequency $f_1 = 5$ Hz, and making decisions at frequency of 1 Hz.
The FoVs are squares with edge lengths of $\{20, 25, 22\} $ m, and sensing quality factors of agents $\alpha_i$ are $\{0.1, 0.15, 0.12\}$.
The trajectories of OOIs are designed with step lengths obeying the Levy walk in the speed interval of [1.0, 3.0] m/s, describing human gaits.
Such curves in Fig. \ref{fig-simenv} contain the general tracking challenges such as crossover, non-linearity and divergent trajectories, which requires sufficient coordination to maintain good tracking performance.
Each agent has the speed limit of $v_{max} = 5$ m/s.
The Optimal Subpattern Assignment (OSPA) \cite{schuhmacher2008consistent} metric is applied as the evaluation of multi-target tracking performance.
For two finite sets $X = \{x_1, ..., x_a\}$ and $Y = \{y_1, ..., y_b\}$ without loss of generality, $b \geq a$, let $\Pi_k$ be the set of permutation on $\{1, 2,..., k\}$. 
The OSPA metrics for these two sets is 
\begin{equation}
    \label{ospa_metric}
    \Bar{d}^{(c)}_p(X, Y) = \biggl[\frac{1}{b} 
    \biggl(\min_{\pi \in \Pi_b} \sum_{i=1}^a d^{(c)}(x_i, y_{\pi(i)}) + c^p(b -a)\biggl)
    \biggl]^\frac{1}{p}
\end{equation}
with $d^{(c)}(x, y) := \min(c, d(x, y))$ for metrics $d$.
In target tracking evaluation, we can regard tracking set $\Theta_k$ and OOI true state $\chi_k$ as the sets $X, Y$.
In \eqref{ospa_metric}, the first part addresses the tracking error with metrics $d$, which is Euclidean distance in our problem; second part is the penalty of cardinally, either missing or false alarm in target tracking.
The parameters $c = 50 \text{ m}, p = 2$ are selected in OSPA, with $c$ the cut-off value of target tracking error, and $p=2$ returns L-2 norm.

\begin{table}[b!]
\centering
\begin{tabular}{c|c|cccc}
\hline 
 $\lambda$ & $H$ & SMA-NBO & MWTP & MCR & Dec-POMDP \\
\hline
   
 \multirow{3}{*}{$\displaystyle 15$} & $\displaystyle 1$ & 616 & 643 & 1517 & - \\
\cline{2-6} 
   & $\displaystyle 3$ & 1401 & 1457 & 3309 & 2192   \\
\cline{2-6} 
   & $\displaystyle 5$ & 2094 & 2033 & 5160 & 2641 \\
\hline 
 \multirow{3}{*}{$\displaystyle 45$} & $\displaystyle 1$ & 600 & 628 & 1639 & -  \\
\cline{2-6} 
   & $\displaystyle 3$ & 1386 & 1393 & 3673 & -  \\
\cline{2-6} 
   & $\displaystyle 5$ & 2028 & 2023 & 5883 & -  \\
\hline 
 \multirow{3}{*}{$\displaystyle 75$} & $\displaystyle 1$ & 574 & 604 & 1662 & - \\
\cline{2-6} 
   & $\displaystyle 3$ & 1307 & 1319 & 3822 & 1789  \\
\cline{2-6} 
   & $\displaystyle 5$ & 1914 & 1924 & 6291 & 2551 \\
 \hline
\end{tabular}
\caption{Average runtime per simulation trial of different algorithms: truncated SMA-NBO (SMA-NBO), SMA-NBO with MWTP (MWTP), Monte Carlo Multi-agent Rollout (MCR) and Dec-POMDP, unit in second.}
\label{table-runtime}
\end{table}

\subsection{A Random Occlusion Forest Environment}

One interest of SMA-NBO algorithm is to achieve non-myopic performance indicated by \eqref{J_H}, which provides the ability to overcome or compensate for occluded areas during tracking.
Also, the behavior of exchanging duty of tracking occluded targets between agents demonstrates fleet coordination.
However, there are few studies in target tracking of the performance evaluation with random occlusion area.
Inspired by studies about collision avoidance in a random obstacle environment \cite{karaman2012high}, we create a random forest of occlusion objects shown in Fig. \ref{fig-simenv}.
Imagine trees of fixed radius generate shadow areas in Fig. \ref{fig-scenario}a, the naive but direct factors impacting the tracking performance are tree density $\lambda$ and shadow radius $R$.
We choose three densities, and the expected numbers of trees in this $150\times 100$ m AOI are $\lambda = \{15, 45, 75\}$, which follow Poisson distribution in the random forest \cite{karaman2012high};
two sizes of tree are selected $R= \{1, 5\}$ m.
For each combination of $(\lambda, R)$, we generate 50 maps that randomizes tree positions without overlapping.
These maps are utilized for statistical performance evaluation of algorithms in the following report.

\subsection{SMA-NBO Tracking Performance}

\begin{figure*}[t!]
    \centering
    \begin{tikzpicture}
        \begin{groupplot}[
            group style={
                group size=3 by 1,
                horizontal sep=5mm,
                y descriptions at=edge left,
            },
            width = 0.33\textwidth,
            height = 0.22\textwidth,
            xmin= 0.1,xmax= 20,
            ymin= 0,ymax= 1.1,
            grid=both,
            grid style={line width=.1pt, draw=gray!10},
            major grid style={line width=.2pt,draw=gray!50},
            ytick distance=0.2,
            xmode=log,
            log ticks with fixed point,
            ylabel={Accumulated Frequency},
            samples=3,
            cycle list name=my cycle list,
            /tikz/smooth,
            /tikz/mark size=0.8,
        ]
        \nextgroupplot[title={$\lambda=15$}]
            \addplot table [mark=none, x=x, y=nbo_0.001_1_1_wtp_False, col sep=comma] {data/merge.csv};
            \addplot table [mark=none, x=x, y=nbo_0.001_5_1_wtp_False, col sep=comma] {data/merge.csv};
            
            \addplot table [mark=none, x=x, y=nbo_0.001_1_3_wtp_False, col sep=comma] {data/merge.csv};
            \addplot table [mark=none, x=x, y=nbo_0.001_5_3_wtp_False, col sep=comma] {data/merge.csv};
            
            \addplot table [mark=none, x=x, y=nbo_0.001_1_5_wtp_False, col sep=comma] {data/merge.csv};
            \addplot table [mark=none, x=x, y=nbo_0.001_5_5_wtp_False, col sep=comma] {data/merge.csv};
            
        \nextgroupplot[title ={$\lambda=45$}, xlabel={OSPA metric (m)},]
            \addplot table [mark=none, x=x, y=nbo_0.003_1_1_wtp_False, col sep=comma] {data/merge.csv};
            \addplot table [mark=none, x=x, y=nbo_0.003_5_1_wtp_False, col sep=comma] {data/merge.csv};
            
            \addplot table [mark=none, x=x, y=nbo_0.003_1_3_wtp_False, col sep=comma] {data/merge.csv};
            \addplot table [mark=none, x=x, y=nbo_0.003_5_3_wtp_False, col sep=comma] {data/merge.csv};
            
            \addplot table [mark=none, x=x, y=nbo_0.003_1_5_wtp_False, col sep=comma] {data/merge.csv};
            \addplot table [mark=none, x=x, y=nbo_0.003_5_5_wtp_False, col sep=comma] {data/merge.csv};
            
        \nextgroupplot[
            title={$\lambda=75$},
            legend pos=outer north east,
            legend cell align=left,
            legend style={font=\tiny},
        ]
            \addplot table [mark=none, x=x, y=nbo_0.005_1_1_wtp_False, col sep=comma] {data/merge.csv};
            \addlegendentry{$R=1, H=1$}
            \addplot table [mark=none, x=x, y=nbo_0.005_5_1_wtp_False, col sep=comma] {data/merge.csv};
            \addlegendentry{$R=5, H=1$}
            
            \addplot table [mark=none, x=x, y=nbo_0.005_1_3_wtp_False, col sep=comma] {data/merge.csv};
            \addlegendentry{$R=1, H=3$}
            \addplot table [mark=none, x=x, y=nbo_0.005_5_3_wtp_False, col sep=comma] {data/merge.csv};
            \addlegendentry{$R=5, H=3$}
            
            \addplot table [mark=none, x=x, y=nbo_0.005_1_5_wtp_False, col sep=comma] {data/merge.csv};
            \addlegendentry{$R=1, H=5$}
            \addplot table [mark=none, x=x, y=nbo_0.005_5_5_wtp_False, col sep=comma] {data/merge.csv};
            \addlegendentry{$R=5, H=5$}
        \end{groupplot}
    \end{tikzpicture}
    \caption{Tracking performance of truncated SMA-NBO in 50 random maps of different density.}
        \label{fig-nbo}
\end{figure*}
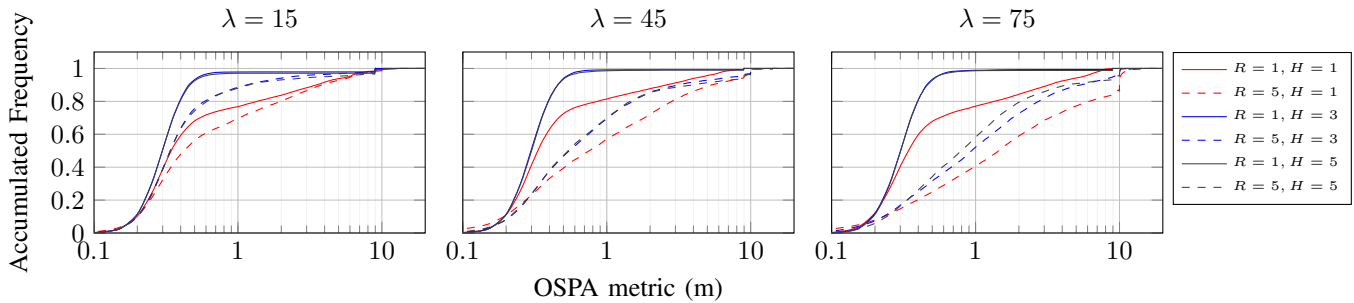

First, the tracking performance of the proposed SMA-NBO is explored in all density and radius random forests; horizon length $H$ is varied since it plays an important role in featuring the coordinated and non-myopic behavior of the agents, with the trade-off in computation.
Longer horizons $H$ avoid greedy behavior in general, specifically when targets are in occlusion areas, and enables agents to realize the targets will exit the occlusion in the future.
Fig. \ref{fig-nbo} reports the tracking performance in the format of accumulated frequency (empirical cumulative distribution function, ECDF) plots over OSPA values, resulting from the simulations over 50 random forests.
It should be noted that same batch of random maps are used in testing different algorithms in same parameter specifications $(\lambda, R)$.
Examination of Fig. \ref{fig-nbo} shows the benefits of longer planning horizons in higher density forests with larger occlusions. 
SMA-NBO achieves 95\%+ OSPA values lower than 1 m in small occlusion size, $R=1$, with horizon $H>1$; in $R=1$, SMA-NBO with $H=5$ reaches 90\%, 70\%, and 60\% OSPA values lower than 1 m in all densities, showing promising tracking performance of SMA-NBO.

Larger horizons in SMA-NBO display benefits that are visible from the frequency plot. 
In cases of $R=1$, i.e., small occlusions, the performance of $H=3$ and $H=5$ is almost identical;
while in the case of $R=5, \lambda=75$, longer horizons highlight performance improvement when compensating dense occlusion coverage in the forest ($39\%$). 
Although density has some impact on tracking performance, the advantages of horizon length are tightly linked with occlusion radii.

The runtime statistics are recorded in Table. \ref{table-runtime}.
For SMA-NBO, the general rule is that simulation time relative to planning horizon length due to the extension of action space via time dimension.
For an intuitive illustration of coordination in SMA-NBO, in the \textbf{video attachment} the behaviors different horizons are compared in the map of $R=5, \lambda=45$.
From 17sec-25sec with $H=5$, the red UAV chases the bottom-left target which is in diverging from others, 25sec-35sec the blue and pink UAVs deal with 3 targets with hand-offs of the center targets. 
Such behavior is lacking in $H=1,3$ because of the limit in the horizon; longer horizons implicitly inspire cooperation.

\subsection{HECTG SMA-NBO}

As described in \eqref{infi-obj}, the horizon can be further approximated with a HECTG function, and MWTP is one candidate detailed in Algorithm \ref{alg-mwtp}.
We test this SMA-NBO + MWTP in random forests of $R=5$ with results shown in Fig. \ref{fig-nbo-mc}.
From the simulation result of Fig. \ref{fig-nbo-mc}, the major performance improvement by MWTP can be seen in $H=1$, which helps generating non-myopic behavior with greedy planning cases; in low density forests ($\lambda = 15, 45$) MWTP $H=1$ even achieves close performance to longer horizon $H=3, 5$.
On the other hand, no significant computation load is introduced from the MWTP term by comparing the runtime of SMA-NBO and MWTP in Table. \ref{table-runtime}.
Thus, significant improvement in performance is seen to acknowledge MWTP as a HECTG without detrimental computational effects.

\subsection{Different Approximation Approaches}
\begin{figure*}[t!]
    \centering
    \begin{tikzpicture}
        \begin{groupplot}[
            group style={
                group size=3 by 3,
                horizontal sep=5mm,
                y descriptions at=edge left,
            },
            width = 0.33\textwidth,
            height = 0.2\textwidth,
            xmin= 0.1,xmax= 20,
            ymin= 0,ymax= 1.1,
            grid=both,
            grid style={line width=.1pt, draw=gray!10},
            major grid style={line width=.2pt,draw=gray!50},
            ytick distance=0.2,
            xmode=log,
            log ticks with fixed point,
            samples=3,
            cycle list name=another list,
            /tikz/smooth,
            /tikz/mark size=0.8,
        ]
        
        \nextgroupplot[
            title={$H=1, \lambda=15$},
            xticklabels=\empty,
        ]
            \addplot table [mark=none, x=x, y=nbo_0.001_5_1_wtp_False, col sep=comma] {data/merge.csv};
            \addplot table [mark=none, x=x, y=nbo_0.001_5_1_wtp_True, col sep=comma] {data/merge.csv};
            \addplot table [mark=none, x=x, y=MonteCarloRollout_0.001_5_1_wtp_False, col sep=comma] {data/merge.csv};
        \nextgroupplot[
            xticklabels=\empty,
            title={$H=3, \lambda=15$}]
            \addplot table [mark=none, x=x, y=nbo_0.001_5_3_wtp_False, col sep=comma] {data/merge.csv};
            \addplot table [mark=none, x=x, y=nbo_0.001_5_3_wtp_True, col sep=comma] {data/merge.csv};
            \addplot table [mark=none, x=x, y=MonteCarloRollout_0.001_5_3_wtp_False, col sep=comma] {data/merge.csv};
            \addplot table [mark=none, x=x, y=decPOMDP_nbo_0.001_5_3_wtp_False, col sep=comma] {data/merge.csv};
            
        \nextgroupplot[
            xticklabels=\empty,
            title={$H=5, \lambda=15$},
            legend pos=outer north east,
            legend cell align=left,
            legend style={font=\tiny}
        ]   
            \addplot table [mark=none, x=x, y=nbo_0.001_5_5_wtp_False, col sep=comma] {data/merge.csv};
            \addlegendentry{SMA-NBO};
            \addplot table [mark=none, x=x, y=nbo_0.001_5_5_wtp_True, col sep=comma] {data/merge.csv};
            \addlegendentry{SMA-NBO + MWTP};
            \addplot table [mark=none, x=x, y=MonteCarloRollout_0.001_5_5_wtp_False, col sep=comma] {data/merge.csv};
            \addlegendentry{MCR};
            \addplot table [mark=none, x=x, y=decPOMDP_nbo_0.001_5_5_wtp_False, col sep=comma] {data/merge.csv};
            \addlegendentry{Dec-POMDP}
        
        \nextgroupplot[
            xticklabels=\empty,
            title={$H=1, \lambda=45$},
            ylabel={Accumulated Frequency},
        ]
            \addplot table [mark=none, x=x, y=nbo_0.003_5_1_wtp_False, col sep=comma] {data/merge.csv};
            \addplot table [mark=none, x=x, y=nbo_0.003_5_1_wtp_True, col sep=comma] {data/merge.csv};
            \addplot table [mark=none, x=x, y=MonteCarloRollout_0.003_5_1_wtp_False, col sep=comma] {data/merge.csv};
        \nextgroupplot[
            xticklabels=\empty,
            title={$H=3, \lambda=45$}]
            \addplot table [mark=none, x=x, y=nbo_0.003_5_3_wtp_False, col sep=comma] {data/merge.csv};
            \addplot table [mark=none, x=x, y=nbo_0.003_5_3_wtp_True, col sep=comma] {data/merge.csv};
            \addplot table [mark=none, x=x, y=MonteCarloRollout_0.003_5_3_wtp_False, col sep=comma] {data/merge.csv};
            
        \nextgroupplot[
            xticklabels=\empty,
            title={$H=5, \lambda=45$},
            legend pos=outer north east,
            legend cell align=left,
            legend style={font=\tiny}
        ]   
            \addplot table [mark=none, x=x, y=nbo_0.003_5_5_wtp_False, col sep=comma] {data/merge.csv};
            \addlegendentry{SMA-NBO};
            \addplot table [mark=none, x=x, y=nbo_0.003_5_5_wtp_True, col sep=comma] {data/merge.csv};
            \addlegendentry{SMA-NBO + MWTP};
            \addplot table [mark=none, x=x, y=MonteCarloRollout_0.003_5_5_wtp_False, col sep=comma] {data/merge.csv};
            \addlegendentry{MCR};
        
        \nextgroupplot[
            title={$H=1, \lambda=75$},
        ]
            \addplot table [mark=none, x=x, y=nbo_0.005_5_1_wtp_False, col sep=comma] {data/merge.csv};
            \addplot table [mark=none, x=x, y=nbo_0.005_5_1_wtp_True, col sep=comma] {data/merge.csv};
            \addplot table [mark=none, x=x, y=MonteCarloRollout_0.005_5_1_wtp_False, col sep=comma] {data/merge.csv};
        \nextgroupplot[
            xlabel={OSPA metric (m)},
            title={$H=3, \lambda=75$}]
            \addplot table [mark=none, x=x, y=nbo_0.005_5_3_wtp_False, col sep=comma] {data/merge.csv};
            \addplot table [mark=none, x=x, y=nbo_0.005_5_3_wtp_True, col sep=comma] {data/merge.csv};
            \addplot table [mark=none, x=x, y=MonteCarloRollout_0.005_5_3_wtp_False, col sep=comma] {data/merge.csv};
            \addplot table [mark=none, x=x, y=decPOMDP_nbo_0.005_5_3_wtp_False, col sep=comma] {data/merge.csv};
            
        \nextgroupplot[
            title={$H=5, \lambda=75$},
            legend pos=outer north east,
            legend cell align=left,
            legend style={font=\tiny},
        ]   
            \addplot table [mark=none, x=x, y=nbo_0.005_5_5_wtp_False, col sep=comma] {data/merge.csv};
            \addlegendentry{SMA-NBO};
            \addplot table [mark=none, x=x, y=nbo_0.005_5_5_wtp_True, col sep=comma] {data/merge.csv};
            \addlegendentry{SMA-NBO + MWTP};
            \addplot table [mark=none, x=x, y=MonteCarloRollout_0.005_5_5_wtp_False, col sep=comma] {data/merge.csv};
            \addlegendentry{MCR};
            \addplot table [mark=none, x=x, y=decPOMDP_nbo_0.005_5_5_wtp_False, col sep=comma] {data/merge.csv};
            \addlegendentry{Dec-POMDP}
        \end{groupplot}
    \end{tikzpicture}
    \caption{Tracking performance comparison of truncated SMA-NBO, SMA-NBO+MWTP and Monte Carlo Multi-agent Rollout (MCR) and Dec-POMDP in random maps of size $R=5$.}
    \label{fig-nbo-mc}
\end{figure*}
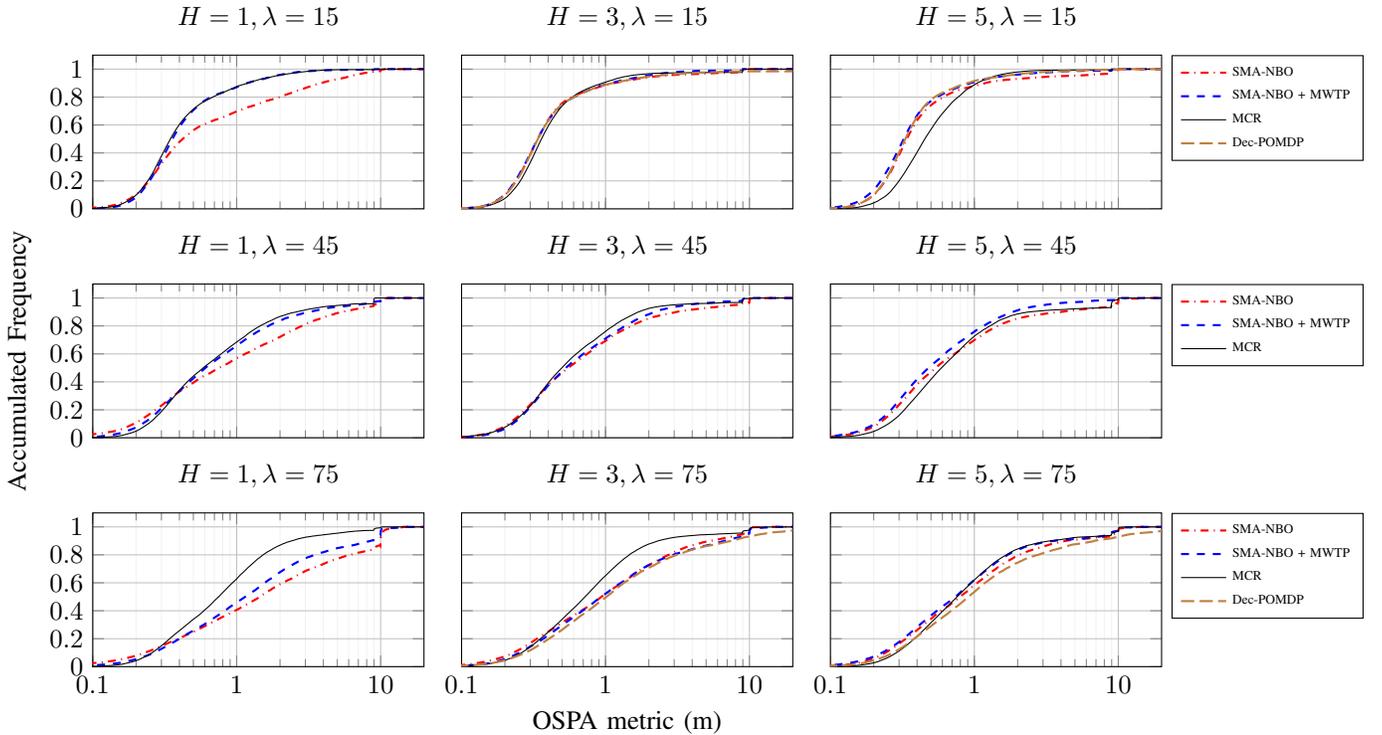

The Monte Carlo method is one of the most popular methods in handling unobservable variables, and it is considered as another approximation approach to future target trajectories \cite{ragi2020random, li2021optimizing}.
Typically it requires a set of samples drawn from the belief distribution $b^{\chi}$, then plans as an optimal control problem accordingly based on the approximate expected objective function with samples.
The number of trajectory samples should be sufficient to obtain a good value estimate.
However, the NBO approach only uses the nominal trajectory as the MAP estimation, which brings an advantage in computation cost.
To study the difference between these two methods, we compared the SMA-NBO with Monte Carlo multi-agent Rollout (MCR) method \cite{li2021optimizing} in random forests of $R=5$.
The latter runs with 50 Monte Carlo trajectory samples in simulation.
From the performance perspective in Fig. \ref{fig-nbo-mc}, MCR provides better tracking than truncated SMA-NBO in low horizon ($H = 1$), but SMA-NBO with MWTP achieves same level of performance as MCR;
when $H>1$, the advantage of MCR is no longer obvious. 
However, the computation in MCR is much more intense reflected by Table. \ref{table-runtime}, which requires over 2.2 times runtime than SMA-NBO.

\subsection{Comparison of Decision-Making Architecture}

Dec-POMDP is a decentralized approach with no decision communication in the fleet.
In Fig. \ref{fig-nbo-mc}, we selectively pick the scenarios of horizon length $H=3, 5$ and the extreme densities $\lambda = 15, 75$ for the comparison of Dec-POMDP (brown lines) and sequential algorithms.
It is worth mentioning that NBO is also applied to solve Dec-POMDP, similar to \cite{ragi2013uav}.
In all four scenarios, there is no significant difference in OSPA distribution between Dec-POMDP and SMA-NBO series algorithms.
Even though each agent in Dec-POMDP generates the fleet-wise optimal planning, the sequential method of SMA-NBO can also achieve similar tracking performance.
On the other hand, from Table \ref{table-runtime}, the runtime difference between SMA-NBO and Dec-POMDP shows the computational advantage of sequential decision making and leveraging the policy of intent.
Given the time complexity's exponential increase by agent number $n$ in Dec-POMDP, the runtime difference will be even larger when more UAVs are in the system.

\section{Conclusion}

This paper introduces the SMA-NBO algorithm as the information-driven planning algorithm of multiple mobile sensors in the task of target tracking. 
Nominal belief-state optimization is adapted for sequential multi-agent decision making, admitting a distributed system architecture.
Specifically, SMA-NBO recycles optimized single-agent policies from the prior decision epoch, constructing a \textit{policy of intent} to inform future agent action selections.
Additionally, this sequential methodology retains tracking performance and reduces the computational load when compared to contemporary distributed methods, e.g., Dec-POMDP and MCR.

We evaluate SMA-NBO, Dec-POMDP and MCR performance in a random occlusion forest parameterized by occlusion size $R$ and density $\lambda$.  
This environment exemplifies the impacts of look-ahead horizon lengths, showing strong correlation to the occlusion size $R$. 
Realizing the benefit of non-myopic decisions, we augment SMA-NBO with a HECTG, namely an adapted MWTP \cite{miller2009coordinated}.  
The statistical result shows SMA-NBO is capable of multi-sensor coordination to track targets in significantly occluded environments.

Based on our investigation of SMA-NBO, the horizon selection of SMA-NBO based on the size and density of the occlusions leads to practical implementation.
Also, theoretic performance boundary of the sequential information-driven planning is worth studying.



\bibliography{IEEEabrv, reference}

\begin{thebibliography}{10}
\providecommand{\url}[1]{#1}
\csname url@samestyle\endcsname
\providecommand{\newblock}{\relax}
\providecommand{\bibinfo}[2]{#2}
\providecommand{\BIBentrySTDinterwordspacing}{\spaceskip=0pt\relax}
\providecommand{\BIBentryALTinterwordstretchfactor}{4}
\providecommand{\BIBentryALTinterwordspacing}{\spaceskip=\fontdimen2\font plus
\BIBentryALTinterwordstretchfactor\fontdimen3\font minus
  \fontdimen4\font\relax}
\providecommand{\BIBforeignlanguage}[2]{{%
\expandafter\ifx\csname l@#1\endcsname\relax
\typeout{** WARNING: IEEEtran.bst: No hyphenation pattern has been}%
\typeout{** loaded for the language `#1'. Using the pattern for}%
\typeout{** the default language instead.}%
\else
\language=\csname l@#1\endcsname
\fi
#2}}
\providecommand{\BIBdecl}{\relax}
\BIBdecl

\bibitem{ferrari2021information}
S.~Ferrari and T.~A. Wettergren, \emph{Information-Driven Planning and
  Control}.\hskip 1em plus 0.5em minus 0.4em\relax MIT Press, 2021.

\bibitem{paull2012sensor}
L.~Paull, S.~Saeedi, M.~Seto, and H.~Li, ``Sensor-driven online coverage
  planning for autonomous underwater vehicles,'' \emph{{IEEE/ASME} Trans.
  Mechatronics}, vol.~18, no.~6, pp. 1827--1838, 2012.

\bibitem{atanasov2015decentralized}
N.~Atanasov, J.~Le~Ny, K.~Daniilidis, and G.~J. Pappas, ``Decentralized active
  information acquisition: Theory and application to multi-robot slam,'' in
  \emph{Proc. IEEE Int. Conf. Robot. Automat.}, 2015, pp. 4775--4782.

\bibitem{cortes2004coverage}
J.~Cortes, S.~Martinez, T.~Karatas, and F.~Bullo, ``Coverage control for mobile
  sensing networks,'' \emph{IEEE Trans. Robot. Autom.}, vol.~20, no.~2, pp.
  243--255, 2004.

\bibitem{krause2008near}
A.~Krause, A.~Singh, and C.~Guestrin, ``Near-optimal sensor placements in
  gaussian processes: Theory, efficient algorithms and empirical studies.''
  \emph{J. Mach. Learn. Res.}, vol.~9, no.~2, 2008.

\bibitem{ragi2013uav}
S.~Ragi and E.~K. Chong, ``Uav path planning in a dynamic environment via
  partially observable markov decision process,'' \emph{{IEEE} Trans. Aerosp.
  Electron. Syst.}, vol.~49, no.~4, pp. 2397--2412, 2013.

\bibitem{krakow2018simultaneous}
L.~W. Krakow, C.~M. Eaton, and E.~K. Chong, ``Simultaneous non-myopic
  optimization of uav guidance and camera gimbal control for target tracking,''
  in \emph{Proc. IEEE Conf. Control Techn. Appl.}, 2018, pp. 349--354.

\bibitem{ramachandran2021resilience}
R.~K. Ramachandran, N.~Fronda, and G.~Sukhatme, ``Resilience in multi-robot
  multi-target tracking with unknown number of targets through
  reconfiguration,'' \emph{IEEE Control Netw. Syst.}, 2021.

\bibitem{li2021optimizing}
T.~Li, L.~W. Krakow, and S.~Gopalswamy, ``Optimizing consensus-based
  multi-target tracking with multiagent rollout control policies,'' in
  \emph{Proc. IEEE Conf. Control Techn. Appl.}, 2021, pp. 131--137.

\bibitem{ragi2020random}
S.~Ragi and H.~D. Mittelmann, ``Random-sampling monte-carlo tree search methods
  for cost approximation in long-horizon optimal control,'' \emph{{IEEE}
  Control Syst. Lett.}, vol.~5, no.~5, pp. 1759--1764, 2020.

\bibitem{corah2019distributed}
M.~Corah and N.~Michael, ``Distributed matroid-constrained submodular
  maximization for multi-robot exploration: Theory and practice,'' \emph{Auton.
  Robots.}, vol.~43, no.~2, pp. 485--501, 2019.

\bibitem{friedrich2019greedy}
T.~Friedrich, A.~G{\"o}bel, F.~Neumann, F.~Quinzan, and R.~Rothenberger,
  ``Greedy maximization of functions with bounded curvature under partition
  matroid constraints,'' in \emph{Proc. AAAI Conf. Artif. Intell.}, vol.~33,
  no.~01, 2019, pp. 2272--2279.

\bibitem{ragi2014decentralized}
S.~Ragi and E.~K. Chong, ``Decentralized guidance control of uavs with explicit
  optimization of communication,'' \emph{J. Intell. Robot. Syst.}, vol.~73,
  no.~1, pp. 811--822, 2014.

\bibitem{bertsekas2021multiagent}
D.~Bertsekas, ``Multiagent reinforcement learning: Rollout and policy
  iteration,'' \emph{IEEE/CAA J. Autom. Sin.}, vol.~8, no.~2, pp. 249--272,
  2021.

\bibitem{miller2009coordinated}
S.~A. Miller, Z.~A. Harris, and E.~K. Chong, ``Coordinated guidance of
  autonomous uavs via nominal belief-state optimization,'' in \emph{Proc. Amer.
  Control Conf.}, 2009, pp. 2811--2818.

\bibitem{battistelli2014kullback}
G.~Battistelli and L.~Chisci, ``Kullback--leibler average, consensus on
  probability densities, and distributed state estimation with guaranteed
  stability,'' \emph{Automatica}, vol.~50, no.~3, pp. 707--718, 2014.

\bibitem{kaelbling1998planning}
L.~P. Kaelbling, M.~L. Littman, and A.~R. Cassandra, ``Planning and acting in
  partially observable stochastic domains,'' \emph{Artif. intell.}, vol. 101,
  no. 1-2, pp. 99--134, 1998.

\bibitem{schuhmacher2008consistent}
D.~Schuhmacher, B.-T. Vo, and B.-N. Vo, ``A consistent metric for performance
  evaluation of multi-object filters,'' \emph{{IEEE} Trans. Signal Process.},
  vol.~56, no.~8, pp. 3447--3457, 2008.

\bibitem{karaman2012high}
S.~Karaman and E.~Frazzoli, ``High-speed flight in an ergodic forest,'' in
  \emph{Proc. IEEE Int. Conf. Robot. Automat.}, 2012, pp. 2899--2906.

\end{thebibliography}

\end{document}